# A novel fusion reactor with chain reactions for proton-boron11

## Shalom Eliezer and Jose M. Martinez-Val

## Institute of Nuclear Fusion, Polytechnic University of Madrid, Madrid, Spain

## 17 November 2019

## Abstract

Using a combination of laser-plasma interactions and magnetic confinement configurations, a conceptual fusion reactor is proposed in this paper. Our reactor consists of the following: 1) A background plasma of boron11 and hydrogen ions, plus electrons, is generated and kept for a certain time, with densities of the order of a milligram/cm$^3$and temperatures of tens of eV. Both the radiation level and the plasma thermal pressure are thus very low. 2) A plasma channel is induced in a solid target by irradiation with a high power laser that creates a very intense shock wave. This mechanism conveys the acceleration of protons in the laser direction. The mechanisms must be tuned for the protons to reach a kinetic energy of 600 keV (note that this value is not a temperature). 3) Those ultra-fast protons enter the background plasma and collide with a boron11 to produce 3 alphas of 2.9 MeV. 4) Fusion born alphas collide with protons of the plasma and accelerate them up to 600 keV causing a chain reaction. 5) A combination of an induction current and a magnetic bottle keeps the chain reaction process going on, for a pulse long enough to get a high energy gain. 6) Materials for the background plasma and the laser target must be replaced for starting a new chain reaction cycle.

## 1. Introduction

Nuclear fusion is the origin of energy in our sun and in the stars. The nuclear fusion in our sun (e.g. 4 protons combine into an alpha) is a weak interaction. Thus the sun's hot plasma confinement is made possible due to the gravitational force of the sun's large mass. In a terrestrial laboratory, between the possible existing reactions, the first choice for controlled fusion energy is the deuterium tritium strong (DT) interaction creating an alpha and a neutron with the release of 17.6 MeV per reaction. The DT has been chosen since the cross section and the rate of this process are the largest for the lowest practical temperatures, of the order of 10 keV. The problem with the DT reaction is that it produces the undesired neutron that can activate radioactively materials.

The cleanest fusion reaction that avoids the neutron problem is the fusion of protons with $^{11}$B (pB11) that creates three alphas (Hora et al, 2017 and references therein). Using lasers, the first p-$^{11}$B 1000 reactions, just above the level of sensitivity were measured (Belyaev et al, 2005). A combination of highly intense proton beams, of energies above MeV produced by picosecond laser pulses intercepting a plasma created by a second irradiated laser beam produced more than one million pB11 reactions (Labaune et al, 2013). At Prague PALS facility the few hundred joules-nanosecond time duration iodine laser interacting with targets containing high boron concentration doped in silicon crystals produced one billion alpha particles (Picciotto et al, 2014) and in an ELI meeting more than $10^{11}$ alphas were reported per laser shot (L. Giuffrida, 2019).

The main problem in solving the energy problem with fusion on our planet for mankind is the difficulty to create it in a controllable and economical way with a positive energy balance. Two different distinctive schemes have been investigated in the past 60 years: (1) Magnetic confinement fusion (MCF) based on high intensity magnetic fields (several teslas) confining low-density ($10^{14}$ cm$^3$) and high temperatures (~10keV) plasmas for long or practically continuous times. (2) Inertial confinement fusion (ICF) based on rapid heating and compressing the fusion fuel to very large densities (Nuckolls et al, 1972) and very high temperatures, larger than 5 keV for the DT fusion reaction and about 600keV for the pB11 fusion. In order to ignite the fuel with less energy it was suggested (Basov et al, 1992; Tabak et al, 1994) to separate the drivers that compress and ignite the target. First the fuel is compressed, then a second driver ignites a small part of the fuel while the created alpha particles heat the rest of the target. This idea is called fast ignition (FI). The fast ignition problem is that the laser pulse does not reach directly the compressed target; therefore many schemes have been suggested (Guskov, 2013) including proton boron fusion (Martinez Val et al, 1996: Eliezer and Martinez Val, 1998).

The novel scheme described here can be used for a combination such as (Eliezer and Mima, 2009) helium3-deuterium (He3-D), deuterium-litium6 (D-$^6$Li), proton-litium6 (p-$^6$Li), proton-litium7 ((p-$^7$Li), etc. In this paper we suggest the clean (i.e. without neutrons) proton-boron11 fusion yielding 3α,

(1) $$p + {}^{11}B \rightarrow 3\,{}^4He + 8.9MeV$$

This new approach to fusion is given schematically in figure 1. Our reactor consists a background plasma with densities of the order of a milligram/cm$^3$ of boron11 and hydrogen ions. A plasma channel or a solid target are irradiated by a high power laser that creates a shock wave containing proton particles with a flow energy of 600 keV that enter the background plasma. Fusion boron alphas collide with protons of the plasma and accelerate them up to 600 keV causing a chain reaction. The number of the alpha particles $N_\alpha$ created in this process is given by

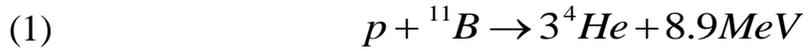

$$N_\alpha \sim 3N_{\alpha 0}\left(e^{\tau/\tau_A} - 1\right)$$

(2)

$$\tau_A \equiv \frac{1}{n_0 \sigma u}$$

$N_\alpha$ is the number of the alpha particles created during the chain reaction for an initial number of alphas $N_{\alpha 0}$. $\tau_A$ is defined as the chain reaction time (also defined as avalanche time, Hora et al, 2017; Eliezer et al, 2016). In our background plasma $n_0$ is of the order of $10^{19}$ cm$^{-3}$, σ ~ 1 barn and u ~ $10^9$ cm/s implying a time $\tau_A \sim 10^{-4}$ s. During an interaction time $\tau \sim 1$ ms we get an increase factor to the originally produced alphas by a factor of the order of $10^4$. In order to keep the chain reaction going our reactor contains a combination of an external electric field and a magnetic mirror confinement device for a pulse long enough to get a high energy gain.

In section 3, the confinement and the chain reactions are described. The conceptual design of the fusion reactor is given in section 4. We end section 5 with a short conclusion.

## 2   The Shock Wave initiating mechanism.

In this system two (or more) shock waves are created by very high irradiance lasers. The desired shock waves are semi-relativistic with a shock wave velocity of the order of 0.1c, where c is the speed of light. The formalism of these shock waves was recently described in the literature (Eliezer et al, 2014; Eliezer et al, 2017). *The laser induced shock wave acts as an accelerator* that accelerates fluid particles inside the shock wave domain (see figures 2) with proton and boron number densities $n_p$ and $n_B$ in a volume V. In our scheme There are two distinct possibilities to accelerate the shocked matter to velocities of the order of $10^9$ cm/s where the center of mass of p-B11 energy of about 600keV, so that the fusion cross section gets its maximum value of $\sigma_{PB11}=1.2\times10^{-24}$ cm$^2$ and the rate of fusion $<\sigma v>$ is about $10^{-15}$ cm$^3$/s. One case the laser irradiates a solid that contains hydrogen and boron (Picciotto et al, 2014; Giuffrida, 2018) in the entrance channel of figure 1, or the other case is the laser creates a shock wave in the background gas.

The physics of shock waves is excellently summarized in Zeldovich and Raizer's book *Physics of Shock Waves and High Temperature Hydrodynamic Phenomena* (Zeldovich and Raizer, 1966). The interaction of a high power laser with a planar target creates a one dimensional (1D) shock wave (Fortov and Lomonosov, 2010; Eliezer, 2013). The theoretical basis for laser induced shock waves analyzed and measured experimentally so far is based on plasma ablation. For laser intensities $10^{12}$ W/cm$^2 < I_L < 10^{16}$ W/cm$^2$ and nanoseconds pulse duration a hot plasma is created. This plasma exerts a high pressure on the surrounding material, leading to the formation of an intense shock wave moving into the interior of the target (Eliezer, 2002). In this paper we are interested in the semi-relativistic shock waves for solid or gas densities. Shock waves induced by lasers with irradiances in this regime are described by relativistic hydrodynamics (Landau and Lifshitz, 1987). Relativistic shock waves were first analyzed by Taub (Taub, 1948) and in the context of laser plasma interactions by Eliezer et. al. (Eliezer et al, 2014).

In the following we use the shock waves equations relevant to figure 1. The relativistic shock wave Hugoniot equations in the laboratory frame of reference are given by

$$(i) \quad \frac{u_p}{c} = \sqrt{\frac{(P_1 - P_0)(e_1 - e_0)}{(e_0 + P_1)(e_1 + P_0)}}$$

$$(3) \qquad (ii) \quad \frac{u_s}{c} = \sqrt{\frac{(P_1 - P_0)(e_1 + P_0)}{(e_1 - e_0)(e_0 + P_1)}}$$

$$(iii) \quad \frac{(e_1 + P_1)^2}{\rho_1^2} - \frac{(e_0 + P_0)^2}{\rho_0^2} = (P_1 - P_0)\left[\frac{(e_0 + P_0)}{\rho_0^2} + \frac{(e_1 + P_1)}{\rho_1^2}\right]$$

P, e and $\rho$ are the pressure, energy density and mass density accordingly, the subscripts 0 and 1 denote the domains before and after the shock arrival, $u_s$ is the shock wave velocity and $u_p$ is the particle flow velocity in the laboratory frame of reference and c is the speed of light. We have assumed that in the laboratory the target is initially at rest. The equation of state (EOS) taken here in order to calculate the shock wave parameters is the ideal gas EOS

$$(4) \qquad e_j = \rho_j c^2 + \frac{P_j}{\Gamma - 1}; \text{ j=0,1.}$$

where $\Gamma$ is the specific heat ratio. We have to solve these equations ((3) and (4)) together with the following piston model equation (Esirkepov et al, 2004; Eliezer et al, 2014),

$$(5) \qquad P_1 = \frac{2I_L}{c}\left(\frac{1 - u_p / c}{1 + u_p / c}\right)$$

It is convenient to use the following laser irradiance and pressure dimensionless variables in the solutions of the above equations

(6)
$$\Pi_L \equiv \frac{I_L}{\rho_0 c^3}; \Pi = \frac{P_1}{\rho_0 c^2} \equiv \frac{P}{\rho_0 c^2}$$

In the transition domain between relativistic and non-relativistic shock waves we get the following solutions for the shock wave parameters

(7)
$$\frac{u_p}{c} = \sqrt{\frac{\Pi(2+\Pi)}{(1+\Pi)(\Gamma+1+\Pi)}} \approx 2\frac{(\Pi_L)^{1/4}}{(\Gamma+1)^{3/4}}$$

$$\frac{u_s}{c} = \sqrt{\frac{\Pi(\Gamma+1+\Pi)}{(1+\Pi)(2+\Pi)}} \approx \sqrt{2}\frac{(\Pi_L)^{1/4}}{(\Gamma+1)^{1/4}}$$

$$\Pi = 2\Pi_L\left(\frac{1-u_p/c}{1+u_p/c}\right) \approx 2\sqrt{\frac{\Pi_L}{\Gamma+1}}$$

The compression $\kappa = \rho/\rho_0$ as a function of the dimensionless pressure $\Pi = P/(\rho_0 c^2)$ is given in figure 2a for $\Gamma = 5/3$. In order to see the transition between the relativistic and nonrelativistic approximation, one has to solve the relativistic equations in order to see the transition effects like the one shown in figure 2a. The numerical solutions shown in figures 2b gives the dimensionless shock wave velocity $u_s/c$ and the particle velocity $u_p/c$ in the laboratory frame of reference versus the dimensionless laser irradiance $\Pi_L = I_L/(\rho_0 c^3)$ in the domain $10^{-4} < \Pi_L < 1$. For the practical proposal the inserted table shows numerical values in the area $10^{-4} < \Pi_L < 10^{-2}$.

Furthermore the speed of sound in units of speed of light, $c_S/c$, as a function of the dimensionless laser irradiance $\Pi_L = I_L/(\rho_0 c^3)$ and the rarefaction velocity, $c_{rw}$, are given by

(8)
$$\frac{c_s}{c} = \sqrt{\left(\frac{\partial P}{\partial e}\right)_S} = \left(\frac{\Gamma P}{e+P}\right)^{1/2} = \left[\frac{\Gamma(\Gamma-1)\Pi}{\Gamma\Pi+(\Gamma-1)\kappa}\right]^{1/2}$$

$$c_{rw} = \frac{c_S+u_p}{1+\left(\dfrac{c_S u_p}{c^2}\right)}$$

The time $\tau_{rw}$ that the rarefaction wave reaches the shock front, for the case that the laser pulse duration is $\tau_L$, is

(9)
$$\tau_{rw} = \frac{c_{rw}\tau_L}{c_{rw} - u_s}$$

As a numerical example we take a laser dimensionless irradiance $\Pi_L$ yielding the desired velocities necessary for our scheme (see figures 2)

$$\Pi_L = I_L / (\rho_0 c^3) = 8.3 \times 10^{-4} \Rightarrow$$

(10)
$$\begin{cases}
\beta \equiv u_p / c = 0.035; \rightarrow u_p = 1.05 \times 10^9 [cm/s] = 4.77\left(\dfrac{e^2}{\hbar c}\right) \\
u_s / c = 0.0465; \rightarrow u_s = 1.40 \times 10^9 [cm/s]; \\
c_s / c = 0.0226; \rightarrow c_s = 0.69 \times 10^9 [cm/s]; \\
c_{rw} / c = 0.0573; \rightarrow c_{rw} = 1.72 \times 10^9 [cm/s]; \\
\Pi = P / (\rho_0 c^2) = 1.61 \times 10^{-3} \rightarrow \\
P[bar] = 1.45 \times 10^{12}\left(\dfrac{\rho_0}{1 g/cm^3}\right)
\end{cases}$$

As an example for the shock wave created in the background plasma, relevant for our next section, we use

(11)
$$\rho_0[g/cm^3] = 10^{-3} \Rightarrow \begin{cases}
I_L[W/cm^2] = 2.24 \times 10^{18} \\
P[bar] = 1.45 \times 10^9 = 1.45 Gb \\
\tau_{rw} = 5.3 \tau_L
\end{cases}$$

For a given laser irradiance $I_L$ and energy $W_L$ we estimate now the laser pulse duration $\tau_L$ for our scheme in order to have a reasonable one dimension (1D) shock wave. $I_L$ is a function of the flow velocity $u_p$ (fixed by $\Pi_L = 8.3 \times 10^{-4}$) and the medium density $\rho_0$, implying

$$I_L[W/cm^2] = 2.2 \times 10^{18} \left( \frac{\rho_0}{10^{-3}\, g/cm^3} \right) \Rightarrow$$

$$S[cm^2] = \pi R_L^{\ 2} = 4.5 \times 10^{-7} \left( \frac{10^{-3}\, g/cm^3}{\rho_0} \right) \left( \frac{W_L}{1kJ} \right) \left( \frac{1ns}{\tau_L} \right)$$

(12)

$$R_L[\mu m] = 0.12 \left[ \left( \frac{1g/cm^3}{\rho_0} \right) \left( \frac{W_L}{1kJ} \right) \left( \frac{1ns}{\tau_L} \right) \right]^{0.5}$$

$$R_L[\mu m] \gg u_s \tau_L [\mu m] = 1.40 \times 10^4 \left( \frac{\mu m}{ns} \right) \tau_L \,(ns)$$

To solve equation (12) we substitute the symbol $\gg$ by a factor of 5 equality, namely a laser diameter larger by a factor of 10 relative to the shock wave length during the pulse duration, $u_s \tau_L$, we get

(13)
$$\tau_L[ns] = 1.2 \times 10^{-3} \left[ \left( \frac{10^{-3}\, g/cm^3}{\rho_0} \right) \left( \frac{W_L}{1kJ} \right) \right]^{1/3}$$

Namely, for our scheme where $I_L = 2.2 \times 10^{18}$ W/cm$^2$ we need a laser pulse duration of 1.2 ps if the laser energy is 1kJ.

## 3   The confinement and the chain reactions.

In order to avoid proton and alphas losses to the wall of the vessel we use a magnetic mirror confinement.  For a longitudinal magnetic field inside the vessel, the transverse radius of the fuel container is at least 2R$_\alpha$, where R$_\alpha$ is the alpha Larmor radius R$_\alpha$, is

(14)
$$R_\alpha = \frac{\gamma \beta_\perp M_\alpha c^2}{2eB}; \beta_\perp = \frac{v_\perp}{c}; \beta = \frac{v}{c}; \gamma = \frac{1}{\sqrt{1-\beta^2}}$$

M$_\alpha$ is the alpha rest mass, e is the elementary charge, B is the applied longitudinal magnetic field and v$_+$ is the perpendicular velocity to the magnetic field. In our case M$_\alpha$ is about 4 times the proton mass and its kinetic energy is about 2.9 MeV, implying $\beta = 3.87 \times 10^{-2}$, $\gamma$-1=7.5$\times 10^{-4}$  and for a magnetic field of 25 Tesla we get R$_\alpha$ about 1 cm. The volume is controlled (Eliezer et al, 1987) by its maximum and minimum magnetic fields $B_{max}$ and $B_{min}$ accordingly where one of the important parameters is its mirror ratio $R_m$ defined by

(15)
$$R_m = \frac{B_{max}}{B_{min}}$$

For $R_m=1.5$ one gets a (minimum) vessel volume $V_v$ given by

(16)
$$V_v = \pi R^2 L = 16\pi R_\alpha{}^3$$
$$R = 2R_\alpha ; L = 4R_\alpha$$

Therefore our mirror confinement of alphas require a minimum volume of about 50 cm³. We can increase the volume to sustain an appropriate fusion energy so that the temperature in the vessel is not more than few electron volt.

In our scheme we use chain reaction (see figure 3) as was explained recently in the literature (Eliezer et al, 2016a) and defined there as avalanche. This avalanche published approach was criticized (Shmatov, 2016; Belloni et al, 2018) and defended (Eliezer et al, 2016b). In this section we show how to keep the chain reaction going for a time duration much larger than the laser pulse duration. This is achieved with an external magnetic field and an accelerating electric field (Bracci and Fiorentini, 1982) acting as a cyclotron for protons and alphas. These fields prolong the avalanche process by overcoming the Bethe-Bloch energy loss (Bethe and Ashkin, 1953) of the protons and the alphas confined in the external magnetic field.

The Bethe-Bloch stopping power dT/dx is given by

(17)
$$\frac{dT_A}{dx}\left[erg / cm\right] = -\frac{4\pi Z_A{}^2 Z_B e^4 n_0}{m_e c^2 \beta^2}\left[\ln\left(\frac{2m_e c^2 \beta^2 \gamma^2}{I}\right) - \beta^2\right]$$

$$\beta = \frac{u}{c}; \quad \gamma = \frac{1}{\sqrt{1-\beta^2}} = 1 + \frac{T_A}{M_A c^2};$$

Non relativistic (NR): $T_A = \frac{1}{2} M_A c^2 \beta^2$

The projectile (e.g. proton in our case) with a mass $M_A$ and a charge $Z_A e$ (e is the positive value of the electron charge) dissipates its energy into the medium (i.e. $H_3B$) via interactions with the electrons of the medium. $T_A$ is the kinetic energy of projectile A, $\beta c$ is the projectile velocity ($\beta \gg 1/137$), index B is the medium where its particles have a charge $Z_B e$. The medium density is $n_0$ [atoms/cm$^3$], $m_e$ is the electron mass and I (~10 eV) is a phenomenological constant describing the binding of the electrons to the medium. We write the stopping power in the following practical units

$$(18) \qquad \frac{dT_A}{dx}\left[ eV / cm \right] = -1.65\times10^7 \, Z_A{}^2 Z_B \left( \frac{n_0}{10^{22} cm^{-3}} \right)\left( \frac{0.04}{\beta} \right)^2$$

For our case $Z_A = 1$ (proton), $Z_B = 2$ ($H_3B$), $\beta = 0.035$ (see equation(10)) and for practical purposes it is conceivable to take $n_0 = 10^{19}$ [cm-3], yielding an electric field of E=43 kV/cm.

A pulsed and oscillatory field is preferable because here it is possible to reach higher peak values than in the static field. In particular, one can use the oscillating electric field in conjunction with a magnetic field B at the cyclotron frequency $\omega_c$ given by

$$\omega_c = \frac{Z_A e B}{M_A c} \quad \text{(Gaussian cgs units); } M_A = A M_p$$

$$(19)$$

$$\omega_c [rad / s, proton] = 9.58\times10^8 \left[ \frac{B}{10[Tesla]} \right]$$

The breakdown field $E_{ac}$ for ac field is much higher than in the dc case and it is approximately given by

$$(20) \qquad E_{ac} \approx \frac{m_e \omega_c c}{e} \approx 43 [kV / cm]\left( \frac{B}{25 Tesla} \right)$$

The number density of the produced alpha particles $n_\alpha$ in the nuclear fusion of pB11 is related to the proton number density (see figure 3),

$$(21) \qquad n_p = n_{p0} + \delta n_p ; \delta n_p = n_\alpha / 3$$

The appropriate number densities [cm$^{-3}$] of the boron11 and protons $n_B$, $n_p$ are

$$\varepsilon = \frac{n_B}{n_p}; \quad n_0 = n_B + n_p = \left(\varepsilon + 1\right)n_p$$

(22)
$$n_p = 5.00 \times 10^{19} \left(\frac{\rho}{10^{-3} \, g \, / \, cm^3}\right) \text{ for } \varepsilon = 1/3$$

$$n_B = 51.66 \times 10^{19} \left(\frac{\rho}{10^{-3} \, g \, / \, cm^3}\right) \text{ for } \varepsilon = 1/3$$

We avoid the protons from decelerating with an external electric field (given by equation(20)) in the H$_3$B medium so that the chain reaction yields the number density of the produced alpha particles $n_\alpha$,

$$\frac{dn_\alpha}{dt} = n_B < \sigma v > \left(3n_{p0} + n_\alpha\right)$$

(23)
$$n_\alpha = 3n_{p0}\left[\exp\left(n_B < \sigma v > t\right) - 1\right] \approx 3n_{p0}\left[\exp\left(\frac{t}{\tau_A}\right) - 1\right]$$

$$\tau_A = \frac{1}{n_B < \sigma v >} \approx 4.8 \times 10^{-5} \left(\frac{10^{-3} \, g \, / \, cm^3}{\rho}\right) \text{ for } \varepsilon = 1/3$$

## 4 The New Reactor.

In this paper, we suggest a clean proton-boron11 fusion reactor (Martinez Val and Eliezer, 2019). The concept of this reactor is distinctly different than ICF and MCF reactors. In this scheme the particles are accelerated by a shock wave to velocities of the order of $10^9$ cm/s so that the fusion cross section gets its maximum value.

We consider the fusion $p + ^{11}B \rightarrow 3\,^4He + 8.9 MeV$. Our reactor consists a background plasma with densities of the order of a milligram/cm$^3$ of boron11 and hydrogen ions. A plasma channel or a solid target are irradiated by a high power laser that creates a shock wave containing proton particles with a flow energy of 600 keV that enter the background plasma. Fusion boron alphas collide with protons of the plasma and accelerate them up to 600 keV causing a chain reaction. The new created alphas are confined by external magnetic fields in the mirror vessel.

We use an existing fluid in our vessel filling the circuit shown in figure 1. The $^{11}B_2H_6$ is the most popular and suitable compound with a density of $1.3 \times 10^{-3}$ g/cm$^3$. The fluid can easily be compressed and reach higher densities if required. Therefore for our numerical examples we take $\rho = 0.001$ g/cm$^3$ and $\varepsilon = 1/3$.

For the case of creating a shock wave in the background plasma in figure 1 we use two PW lasers with irradiances of $10^{18}$ W/cm$^2$. The existence of these lasers is thanks to the chirped-pulse amplification technique developed more than 30 years ago (Strickland and Mourou, 1985; Mourou et al, 1998). Today the laser intensity has increased presently to a maximum value of $10^{22}$ W/cm$^2$ at infrared wavelength (1.6 eV). Laser systems with even higher power are worldwide under consideration and development, such as the ELI project in 3 countries (the Czech Republic, Hungary and Rumania), XCELS in Russia, HIPER in the UK and GEKKO EXA in Japan.

The PW lasers create a semi-relativistic shock wave with flow velocities as given in section 2 and figure 2b. The volume of this accelerated fluid is according to equations (10), (11) and (12),

$$V = S\left(\tau_L + \Delta t\right)u_s \approx 5.3\tau_L u_{sL}S$$

(24)
$$V[cm^3] = 2.5 \times 10^{-6}\left(\frac{10^{-3}\,g/cm^3}{\rho_0}\right)\left(\frac{W_L}{1kJ}\right)$$

Using equations (24), (23) and (22) one gets

$$N_{p0} = n_{p0}V = 1.25 \times 10^{14}\left(\frac{W_L}{1kJ}\right),$$

(25)
$$N_\alpha = 3N_{p0}\left[\exp\left(\frac{t}{\tau_A}\right) - 1\right]; \tau_A \approx 4.8 \times 10^{-5}\left(\frac{10^{-3}\,g/cm^3}{\rho}\right)$$

$$t \gg \tau_A : N_\alpha \approx 1.65 \times 10^{14}\left(\frac{W_L}{1kJ}\right)\exp\left(\frac{t}{\tau_A}\right)$$

The energy of the proton-boron11 nuclear fusion of 8.9MeV per reaction is mainly divided between the 3 alphas so that we can define the gain G in this case by the ratio of $W_\alpha N_\alpha$ to the laser energy $W_L$, where $W_\alpha = (8.9\text{MeV}/3)$. For a reactor facility, a gain of 100 would be economically and technologically satisfactory. Using (25) we have

(26)
$$G = \frac{N_\alpha W_\alpha}{W_L} = 7.8 \times 10^{-2} \exp\left(\frac{t}{\tau_A}\right) \geq 100$$

$$\exp\left(\frac{t}{\tau_A}\right) \geq 1282 \rightarrow t[s] \geq 7.15\tau_A = 3.0 \times 10^{-4}\left(\frac{10^{-3} \, g \, / \, cm^3}{\rho}\right)$$

For each laser pulse we need an electric field with a duration of 0.3 ms in order to receive our chain reaction process. For a 100 MW power reactor, we need the power of 100 lasers pulses. This can be accomplished with 100 lasers operating with a frequency of 1Hz or for example with two lasers of 50 Hz.

## 5. Conclusions

Using a combination of laser-plasma interactions and magnetic confinement a conceptual fusion reactor is suggested in this paper. Our reactor consists of a background plasma with densities of the order of a milligram/cm$^3$, of boron11 and hydrogen ions. Since the temperature of this plasma is few eV the well-known problem with a thermal plasma for pB11 is avoided, namely the radiation level is very low. The fusion process is started via a plasma channel or a solid target that are irradiated by a high power laser that creates a semi-relativistic shock wave. This accelerates a proton beam to 600 keV that are entering the background plasma and collide with boron11 to produce 3 alphas of 2.9 MeV. The fusion boron alphas collide with protons of the plasma and accelerate them up to 600 keV causing a chain reaction as described in figure 3. It is important to emphasize that in a thermal fusion reactor the maximum possible theoretical gain is 8900/600~15, however in our case due to the chain reaction process the maximum gain is (8900/600)*(chain reaction factor) where the chain factor is given by exp[t/(n$_0\sigma$u)] (see equations (2) and (25)) and can be very large. In order to achieve this a combination of an external electric field and a magnetic mirror confinement device keep the chain reaction process going on, for a pulse long enough to get a high energy gain.

**Figure Captions**

<u>Figure 1</u>. *The fusion reactor schematic model.* The two injected high power lasers, in the magnetic field's mirror, trigger the pB11 fusion in the vessel. The created alphas heat the target which flows into a cooler that recirculate the plasma fluid with a density of the order of 1 mg/cm$^3$. The deposition of the energy is done continuously from the cooler while the created alphas are confined by an external magnetic field.

<u>Figure 2.</u>(a) The shock wave compression $\kappa = \rho/\rho_0$ as a function of the shock wave dimensionless pressure $\Pi = P/(\rho_0 c^2)$ is presented. The numerical values are obtained for $\Gamma = 5/3$. The semi-relativistic domain is defined on the red line. (b) The dimensionless shock wave velocity $u_s/c$ and the particle velocity $u_p/c$ in the laboratory frame of reference are given versus the dimensionless laser irradiance $\Pi_L = I_L/(\rho_0 c^3)$ in the domain $10^{-4} < \Pi_L < 1$.

<u>Figure 3.</u> The chain reaction process[31] is described. The medium is a fluid proton-boron mixture and the confinement of the charged particles is done by the external mirror magnetic field.

**Figure 1**.

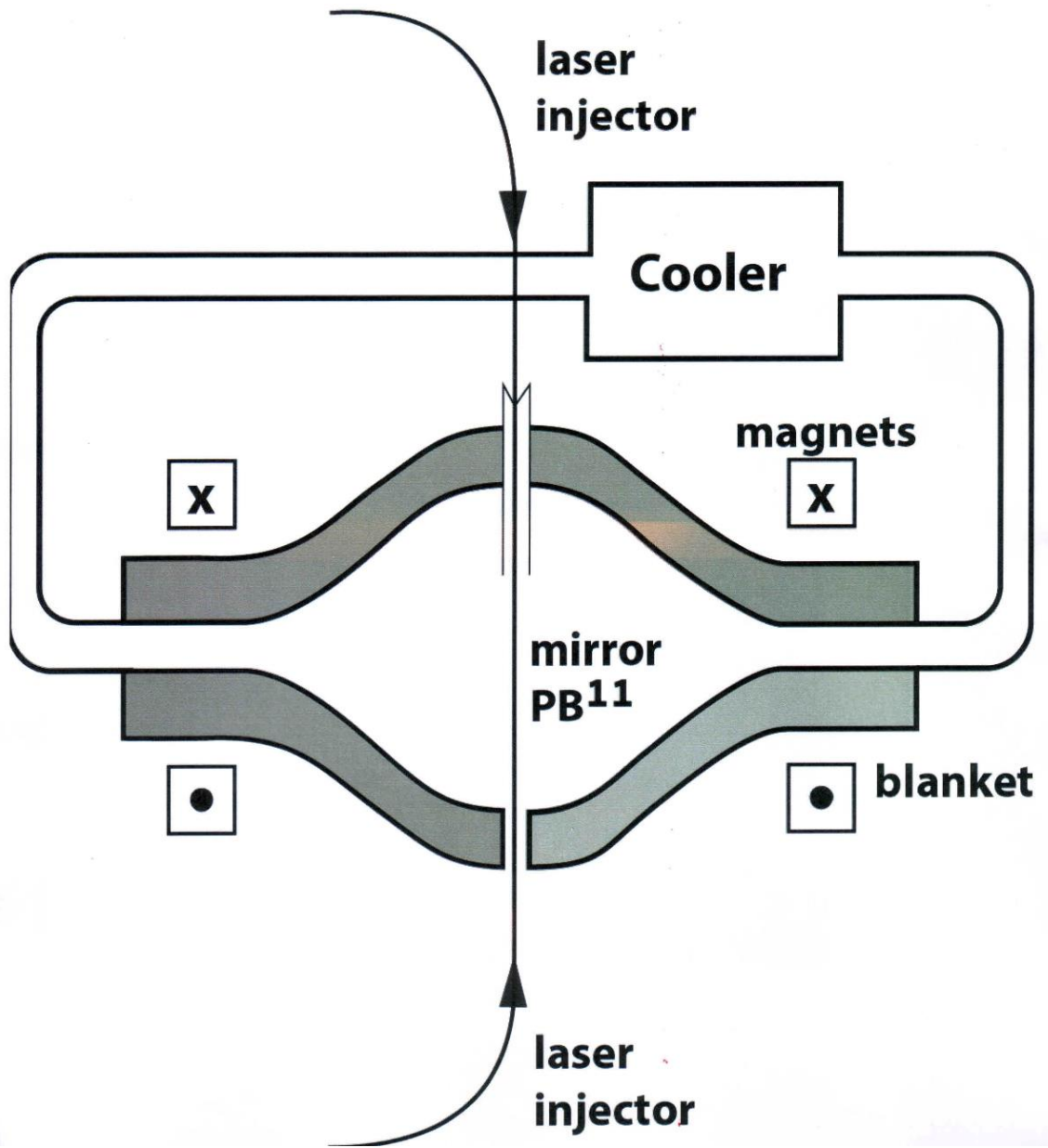

**<u>Figure 2</u>**

(a)

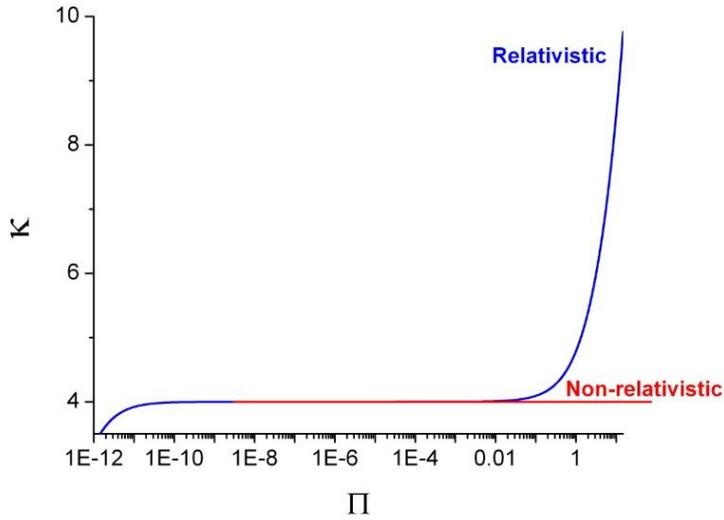

(b)

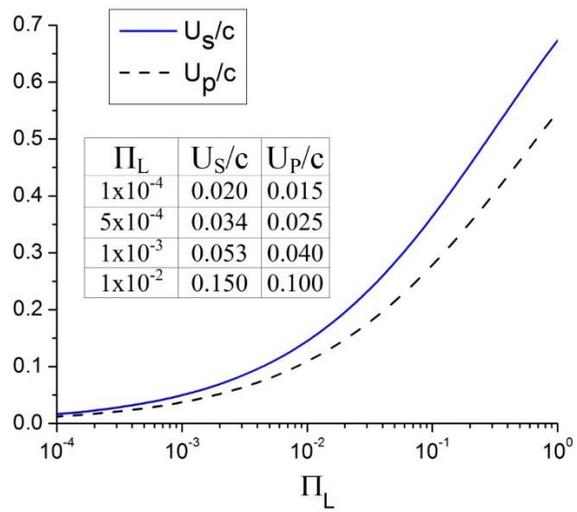

**Figure 3**.

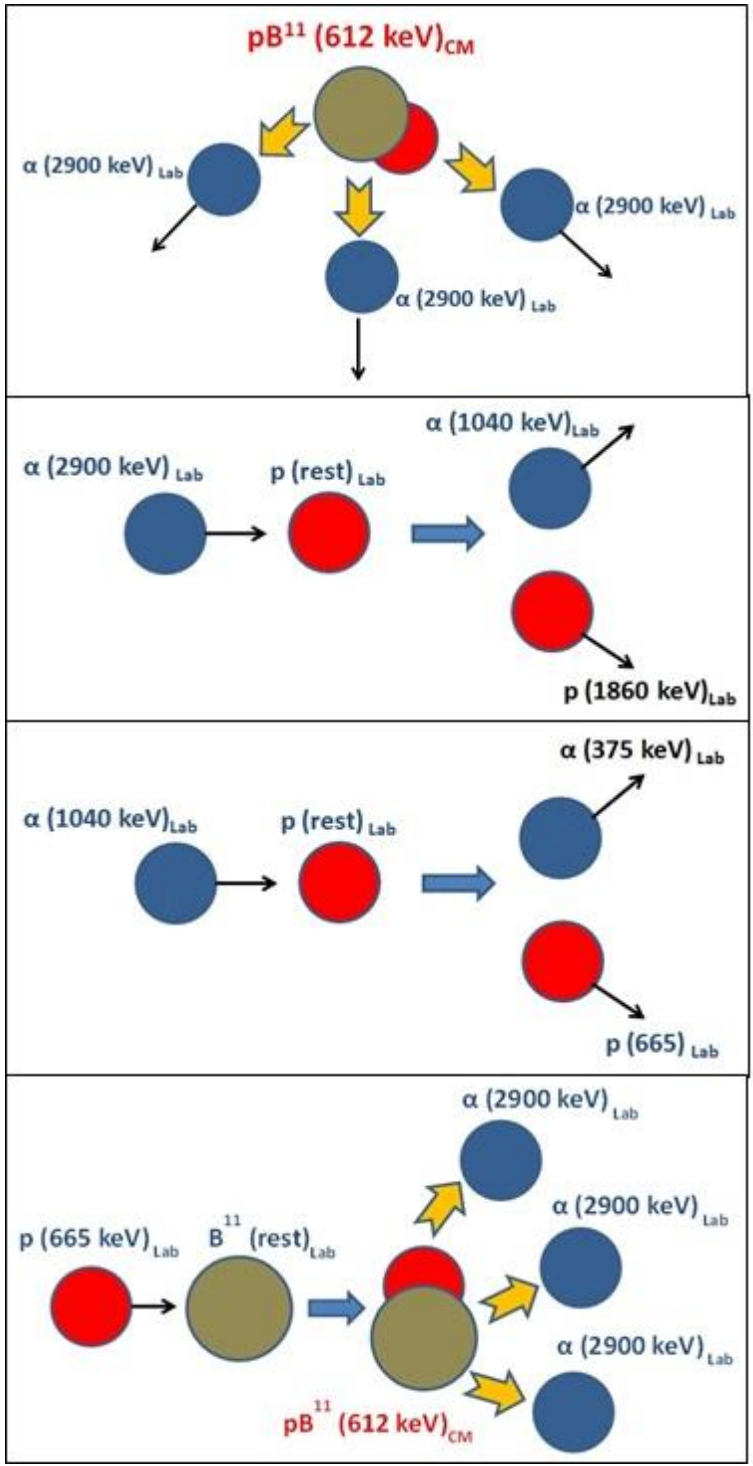